\documentclass[]{jfm}

\usepackage{graphicx}
\usepackage{placeins}
\graphicspath{{./}}
\usepackage{newtxtext}
\usepackage{newtxmath}
\usepackage{amsmath,amssymb}
\usepackage{natbib}
\usepackage{hyperref}
\usepackage[dvipsnames]{xcolor}
\hypersetup{
    colorlinks = true,
    urlcolor   = blue,
    citecolor  = black,
}

\title{The effect of shear-thinning on the scalings and small-scale structures of turbulence}

\author{Marco Edoardo Rosti \corresp{\email{marco.rosti@oist.jp}}}

\affiliation{Complex Fluids and Flows Unit, Okinawa Institute of Science and Technology Graduate University, 1919-1 Tancha, Onna-son, Okinawa 904-0495, Japan}

\begin{document}
\maketitle

\begin{abstract}
We study the homogeneous isotropic turbulence of a shear-thinning fluid modeled by the Carreau model and show how the variable viscosity affects the multiscale behaviour of the turbulent flow. We show that Kolmogorov theory can be extended to such non-Newotnian fluids, provided that the correct choice of average is taken when defining the mean Kolmogorov scale and dissipation rate, to properly capture the effect of the variable viscosity. Thus, the classical phenomenology \textit{à la} Kolmogorov can be observed in the inertial range of scale, with the energy spectra decaying as $k^{-5/3}$ and the third order structure function obeying the $4/5$ law. The changing viscosity instead strongly alters the small scale of turbulence, leading to an enhanced intermittent behavior of the velocity field.
\end{abstract}


\section{Introduction}\label{sec:introduction}
Turbulence is a state of erratic, chaotic, and unpredictable fluid motion at very high Reynolds numbers -- the ratio of the fluid's inertial forces to its viscous forces \citep{frisch_1995k, pope_2001a}. Statistically stationary, homogeneous and isotropic turbulent flows represent the most basic example of the fundamental problem of turbulence, where the statistical characteristics of velocity fluctuations can be probed \citep{ishihara_gotoh_kaneda_2009a}. Typically, energy is injected into the flow at large length scales, whereas viscous processes play a significant role in dissipating energy from the flow at the small scales \citep{alexakis_biferale_2018y}. Scaling exponents are universal in the intermediate range of scales, meaning they are independent of the turbulence generation process, with the energy spectrum decaying as $k^{-5/3}$, as predicted by Kolmogorov's dimensional arguments \citep{kolmogorov_1941a}.

Both in nature and industry, turbulent flows are frequently multi-phase, meaning they are packed with particles, may consist of fluid mixes, or contain additives like polymers, and complex dynamics can arise, such as fluid elasticity \citep{benzi_ching_2018f}, yield stress \citep{balmforth_frigaard_ovarlez_2014a}, or shear-dependent viscosity \citep{larson_desai_2015a}. In recent years, a lot of research has been done on fluids that include long-chain polymers, since adding high molecular weight polymers to a turbulent pipe flow at small concentrations can significantly lower the drag \citep{white_mungal_2008a}. Fluid elasticity can also lead to flow instabilities \citep{shaqfeh_1996a} and turbulence \citep{steinberg_2021p}, even at very small Reynolds numbers. The disordered fluid motion that results from elastic instabilities at low or vanishing Reynolds numbers exhibits characteristics resembling those of classical Newtonian turbulence, such as a power-law scaling in the velocity spectrum and intermittency \citep{datta_ardekani_arratia_beris_bischofberger_mckinley_eggers_lopez-aguilar_fielding_frishman_others_2022a,  singh_perlekar_mitra_rosti_2024a}. At all Reynolds numbers, however, the energy cascade from large to small scales is significantly altered by the presence of polymers, with the turbulent kinetic energy spectrum exhibiting a multiscale behaviour \citep{valente_da-silva_pinho_2014m, rosti_perlekar_mitra_2023a} and the cascade process being altered not only quantitatively but also qualitatively, thus strongly departing from Kolmogorov theory.

A less studied effect of polymers is to provide a shear-dependent viscosity, i.e., with a nonlinear relation between the shear stress and the shear rate, which is often difficult to decouple from the fluid elasticity. When the shear stress increases more than linearly with the shear rate, the fluid is called dilatant or shear-thickening, whereas in the case of opposite behavior, i.e., when the shear stress increases less than linearly with the shear rate, the fluid is called pseudoplastic or shear-thinning. In this work, we focus on fluids whose rheological response does not depend explicitly on time, but only on the present shear rate, often called generalized Newtonian fluids \citep{bird_1976a}. Recent results \citep{rosti_perlekar_mitra_2023a, soligo_rosti_2023a, amor_soligo_mazzino_rosti_2024a} have shown a decay of the energy spectrum of such fluids similar to the one predicted by Kolmogorov for Newtonian fluids, but a clear understanding is still missing, as well as a theory to explain this behaviour, which are the focus of the present work. Thus, after introducing the governing equation, the fluid model, and the setup considered in Section~\ref{sec:methodology}, we analyze the results obtained by direct numerical simulations in Section~\ref{sec:results}, and summarize our findings in Section~\ref{sec:conclusions}.

\section{Methodology}\label{sec:methodology}
We consider an incompressible three-dimensional turbulent flow field, governed by the generalised Navier–Stokes equations. In an inertial, Cartesian frame of reference the equations of momentum and mass conservation for the incompressible flow read as
\begin{equation}
\rho \left( \frac{\partial u_i}{\partial t} + \frac{\partial u_i u_j}{\partial x_j} \right) = -\frac{\partial p}{\partial x_i} + \frac{\partial 2 \mu \mathcal{S}_{ij}}{\partial x_j} + \rho f_i^T \;\;\;\textrm{and}\;\;\; \frac{\partial u_i}{\partial x_i} = 0 \label{eq:NS}
\end{equation}
where $u_i$ is the fluid velocity field, $p$ the pressure, $\mathcal{S}_{ij}=\frac{1}{2}\left( \frac{\partial u_i}{\partial x_j} + \frac{\partial u_j}{\partial x_i} \right)$ the strain rate tensor, $f_i^T$ the force used to sustain the turbulent flow, and $\rho$ and $\mu$ are the density and dynamic viscosity of the fluid (being their ratio $\nu=\mu/\rho$ the kinematic viscosity). Note that, we are using the Einstein sum rule for repeated indices. The forcing  is used to generate and sustain a fully turbulent flow with quasi homogeneous, isotropic, and stationary statistics; in particular, turbulence is sustained using the Arnold-Beltrami-Childress (ABC) cellular-flow forcing, which is a combination of sinusoids with a wavelength equal to the domain size and amplitude $F^T$, kept the same for all cases. The equations of motion are solved in a triperiodic cubic box of size $L$, with periodic boundary conditions applied in all the three Cartesian directions, discretised with $1024$ grid points in each direction to ensure that all the scales down to the smallest dissipative ones are properly solved, i.e., $\eta \lesssim \Delta x$, where $\Delta x$ is the grid spacing and $\eta$ is the Kolmogorov scale of the Newtonian simulation. While this criterion may not be directly applicable to the shear-thinning fluids, we ensure the correctness of the results by using a grid which is suitable for a Reynolds number larger than the one reached here \citep[see e.g.][who reached $Re_\lambda \approx433$]{olivieri_cannon_rosti_2022a, cannon_olivieri_rosti_2024a}.

In the current study, the fluid is non-Newtonian, and we focus on the simple inelastic power-law models, where the local viscosity of the fluid is a function of the sole local value of shear rate, i.e., $\mu=K \dot{\gamma}^{n-1}$, where $n$ is the flow power index and $K$ the fluid consistency index. A Newtonian behavior is recovered when $n=1$, while values of the flow index above and below unity, $n>1$ and $n<1$, denote shear-thickening and shear-thinning fluids, respectively. The consistency index $K$ measures how strong the fluid responds to the imposed deformation rate. 
In the previous relation, the local shear rate $\dot{\gamma}$ is the second invariant of the strain-rate tensor $\mathcal{S}_{ij}$ and is computed by its dyadic product, i.e., $\dot{\gamma}=\sqrt{2 \mathcal{S}_{ij} \mathcal{S}_{ij}}$. The viscosity of a power-law shear-thinning fluid becomes infinite for null shear rate; to overcome this, the Carreau fluid model is used instead, in which the local viscosity is computed as
\begin{equation} \label{eq:carreau}
\mu=\mu_\infty + \left( \mu_0 - \mu_\infty \right) \left[ 1 + \left( \mathcal{K} \dot{\gamma} \right)^2 \right]^\frac{n-1}{2}.
\end{equation}
In this equation, $\mu_0$ and $\mu_\infty$ indicate the lower and upper limits of fluid viscosity at zero and infinite shear rates, and the flow index $n$ and time constant $\mathcal{K}$ have similar interpretations to the material properties of the power-law fluid. 

We use the in-house Fortran code \href{https://www.oist.jp/research/research-units/cffu/code}{\textit{Fujin}} to simulate the flow. The fluid equations are solved numerically on a staggered grid using a second order finite difference code, with pressure points located at the cell center and velocity components at the cell faces. Equations \ref{eq:NS} are advanced in time by a fully explicit fractional step-method, based on the second-order Adams-Bashforth method; finally, the Poisson equation is solved by Fast Fourier Transform.

\subsection{Setup}\label{sec:setup}

\begin{figure}
  \centering
  \includegraphics[width=0.49\textwidth]{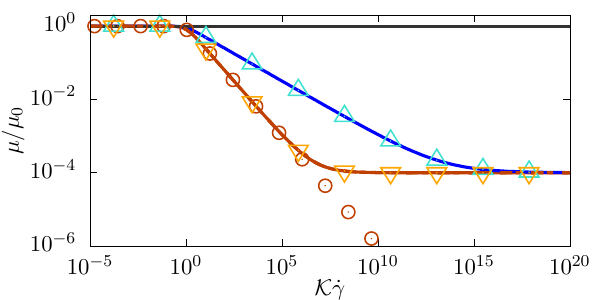}
  \includegraphics[width=0.49\textwidth]{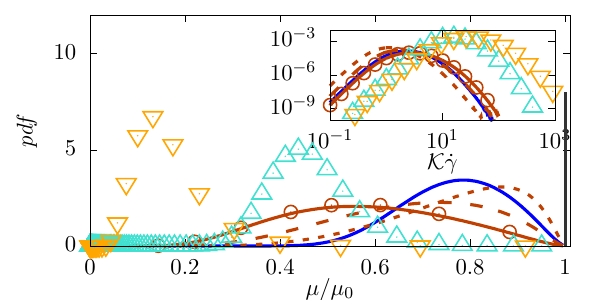}
  \caption{($a$) Fluid viscosity as a function of the shear rate. ($b$) Probability density function of the fluid viscosity (main) and shear rate (inset) in the turbulent flows. The dark gray, blue and brown colors distinguish the different power index $n=1$, $0.7$ and $0.4$, respectively. The solid, dashed and dotted lines distinguish the different Reynolds numbers $Re_\lambda \approx 300$, $200$, and $100$. The circles are the case with $\mu_\infty=0$, while the cyan and gold triangles the case with different $\mathcal{K}$ and $n=0.7$ and $0.4$.}
  \label{fig:visc}
\end{figure}

\begin{table}
\caption{Main input and output parameter of the cases investigated in the present study, together with the color and line styles used throughout the work.}
\label{tab:cases}
\centering
\begin{tabular}{c|cccc|cccc}
~	& $Re_{ABC}$	& $n$	& $\mathcal{K}\sqrt{F^T/L}$	& $\mu_0/\mu_\infty$	& $Re_\lambda$	& $\langle \left( \mu/\mu_0\right) \rangle$	&  $\langle \left(\mu/\mu_0\right)^2 \rangle$	& $\langle \left( \mu/\mu_0\right)^3 \rangle$ \\ \hline 
\textcolor[HTML]{333333}{\huge ---}	&	$7043$	& $1$	&	$-$ 	&	$1$ 	&	$313$	&	1	&	0	&	0	\\
\textcolor[HTML]{333333}{\huge -~-}	&	$3522$	& $1$	&	$-$ 	&	$1$ 	&	$211$	&	1	&	0	&	0	\\
\textcolor[HTML]{333333}{\LARGE \ldots}	&	$1761$	& $1$	&	$-$	& 	$1$ 	&	$138$	&	1	&	0	&	0	\\
\textcolor[HTML]{0000ff}{\huge ---}	&	$7043$	& $0.7$	&	$0.0223$ 	&	$10^4$	&	$323$	&	$0.7492$	&	$0.1720$	&	$-0.0785$	\\
\textcolor[HTML]{0000ff}{\huge -~-}	&	$3522$	& $0.7$	&	$0.0223$ 	&	$10^4$	&	$211$	&	$0.8086$	&	$0.1996$	&	$-0.1011$	\\
\textcolor[HTML]{0000ff}{\LARGE \ldots}	&	$1761$	& $0.7$	&	$0.0223$ 	&	$10^4$	&	$146$	&	$0.8672$	&	$0.2284$	&	$-0.1287$	\\
\textcolor[HTML]{c04000}{\huge ---}	&	$7043$	& $0.4$	&	$0.0223$ 	&	$10^4$	&	$331$	&	$0.5255$	&	$0.0969$	&	$-0.0305$	\\
\textcolor[HTML]{c04000}{\huge -~-}	&	$3522$	& $0.4$	&	$0.0223$ 	&	$10^4$	&	$221$	&	$0.6328$	&	$0.1321$	&	$-0.0487$	\\
\textcolor[HTML]{c04000}{\LARGE \ldots}	&	$1761$	& $0.4$	&	$0.0223$ 	&	$10^4$	&	$143$	&	$0.7415$	&	$0.1732$	&	$-0.0764$	\\
\textcolor[HTML]{c04000}{$\circ$~~$\circ$~~$\circ$}	&	$7043$	& $0.4$	&	$0.0223$ 	&	$\infty$	&	$336$	&	$0.4948$	&	$0.0875$	&	$-0.0264$	\\
\textcolor[HTML]{40e0d0}{$\bigtriangleup ~ \bigtriangleup ~ \bigtriangleup$}	&	$1761$	& $0.7$	&	$0.2230$ 	&	$10^4$	&	$190$	&	$0.4472$	&	$0.0618$	&	$-0.0164$	\\
\textcolor[HTML]{ffa500}{$\bigtriangledown ~ \bigtriangledown ~ \bigtriangledown$}	&	$1761$	& $0.4$	&	$0.2230$ 	&	$10^4$	&	$300$	&	$0.1410$	&	$0.0184$	&	$-0.0010$	\\
\end{tabular}
\end{table}

We consider various fluids with shear dependent viscosity, and report their rheology in Figure~\ref{fig:visc}$a$. In particular, we consider a Newtonian fluid with uniform viscosity $\mu_0$, and two shear-thinning fluids with shear dependent viscosity $\mu$ defined by Equation~\ref{eq:carreau}. These have a fixed ratio of zero to infinity viscosity $\mu_0/\mu_\infty$ equal to $10^4$, and varying power index $n$ equal to $0.7$ and $0.4$. Finally, the consistency index $\mathcal{K}$ is chosen to ensure a low probability of being in the plateaus of viscosity at $\dot{\gamma} \to 0$ and $\infty$. All the fluids are tested in a number of simulations with different values of Reynolds number $Re_{ABC}$, defined based on the amplitude of the forcing and on the zero-shear viscosity $\mu_0$. The parameters are chosen to achieve in the purely Newtonian case a Taylor Reynolds number of $Re_\lambda = \rho u_\textrm{rms} \lambda/\langle \mu \rangle \approx 300$, $200$, and $100$, where $u_\textrm{rms} = \sqrt{\langle u_i u_i \rangle/3}$ is the root mean square of the fluid velocity and $\lambda=u_\textrm{rms} \sqrt{15 \langle \mu \rangle/ \left( \rho \langle \varepsilon \rangle \right) }$ the Taylor lengthscale, with $\langle \varepsilon \rangle = \langle 2 \nu \mathcal{S}_{ij} \mathcal{S}_{ij} \rangle$ the mean turbulent dissipation rate, with $\langle \cdot \rangle$ indicating average over time, space and ensemble. In order to show that the results are general and independent of the value of $\mu_0/\mu_\infty$ and $\mathcal{K}$, we perform three additional simulations, one where we set $\mu_\infty=0$, and another two where we increase $\mathcal{K}$. All the simulations parameter are summarised in Table~\ref{tab:cases}. We allow the flow to reach a statistically steady state, and measure the statistics presented in the rest of the manuscript, averaged in time over $40$ snapshots collected over around $15$ large-eddy turnover times $L/u_{rms}$.

\section{Results}\label{sec:results}
\begin{figure}
  \centering
  \includegraphics[width=0.3\textwidth]{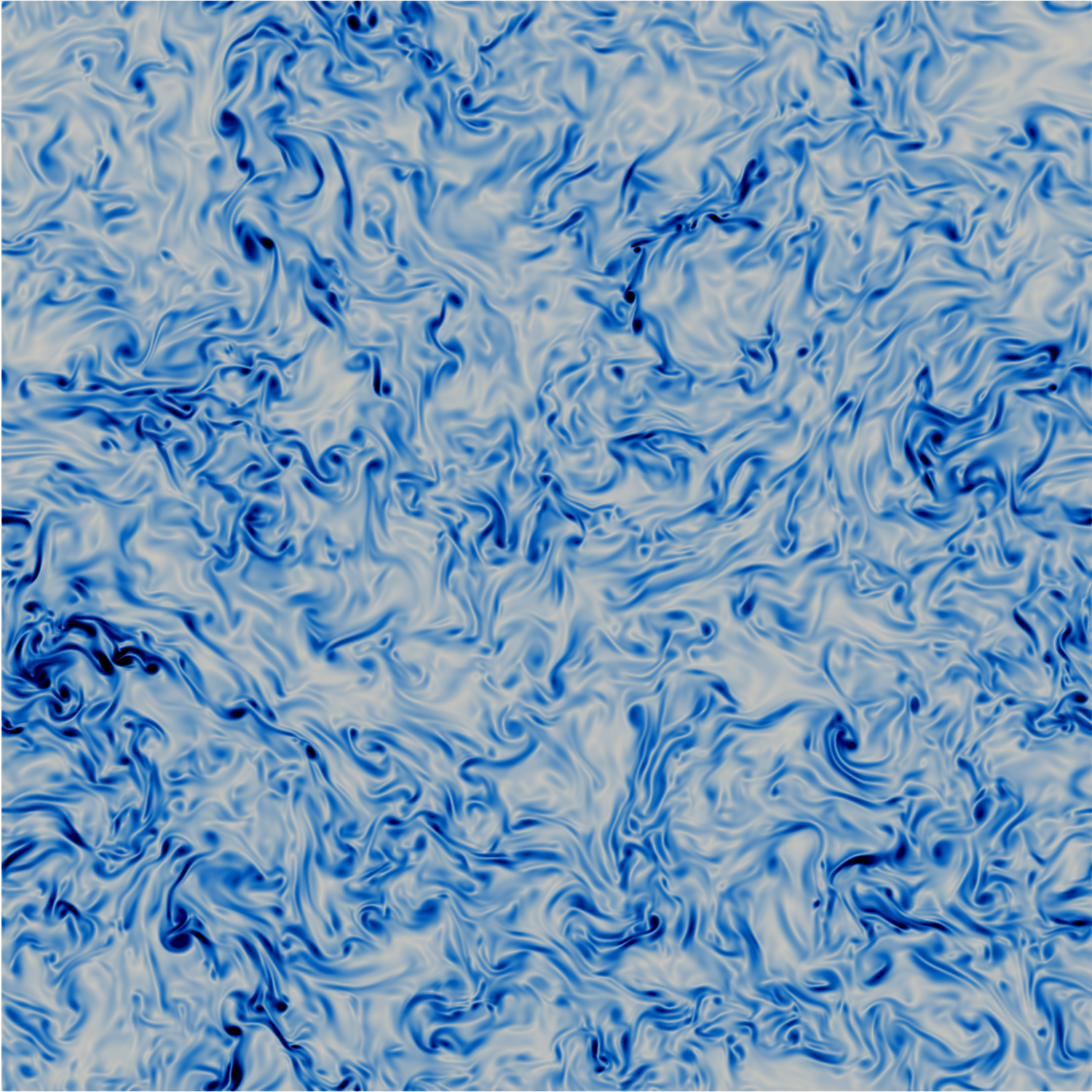}
  \includegraphics[width=0.3\textwidth]{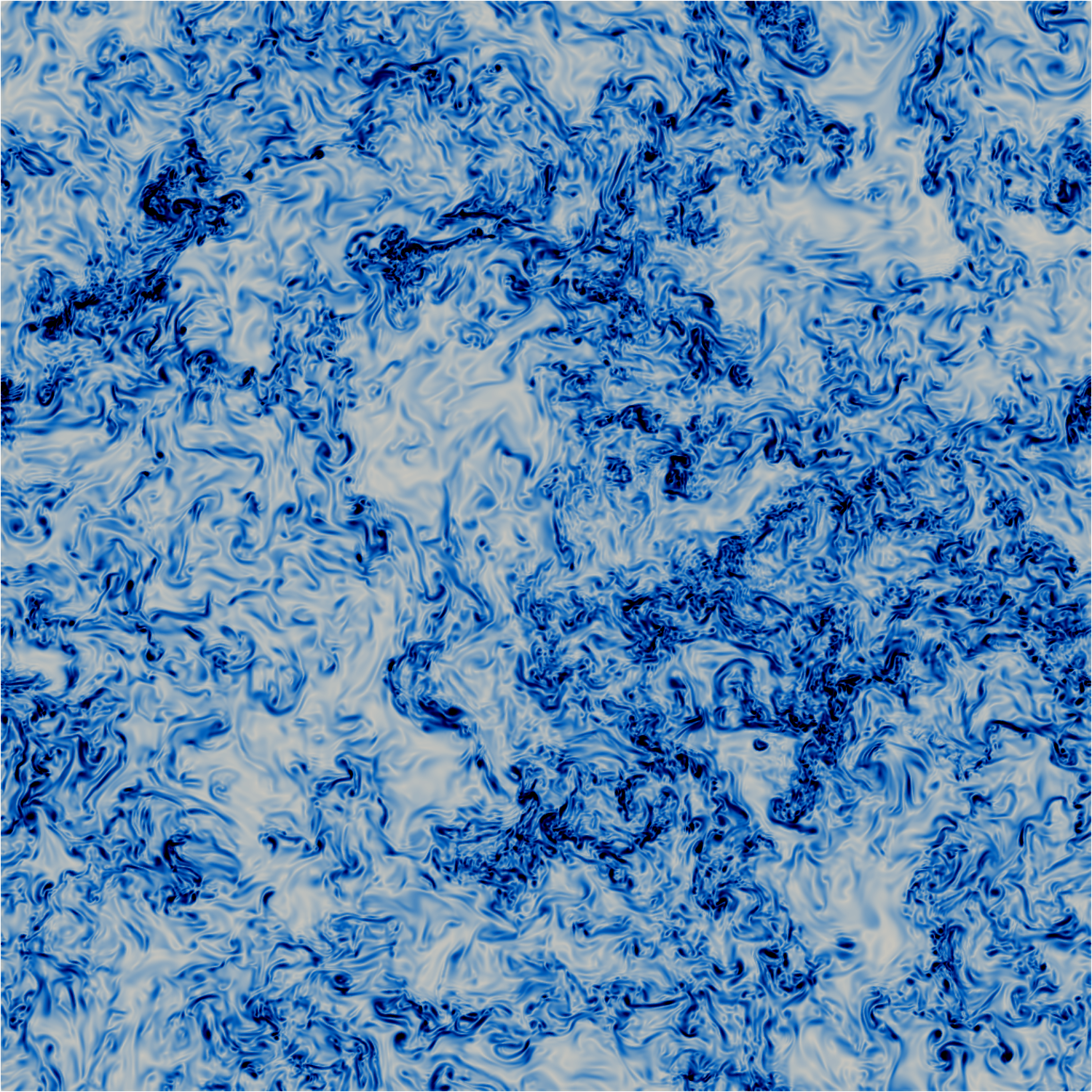}
  \caption{Instantaneous visualization of the vorticity magnitude on a plane in the middle of the cubic box for the (left) Newtonian $n=1$ and (right) shear-thinning fluids with $n=0.4$. Colors go from white to blue from zero to maximum vorticity.}
  \label{fig:visualization}
\end{figure}

We start the analysis of how turbulence is affected by the shear-dependent viscosity by looking at the changes of the Taylor Reynolds number in Table~\ref{tab:cases}. Interestingly, we notice that the Taylor Reynolds number does not significantly change for the different fluids, provided that the averaged viscosity is used in the definitions of the Reynolds number and in the Taylor microscale, while the local viscosity is used when averaging the turbulent dissipation rate. This result is significant, hinting towards how adapting Kolmogorov theory to the generalised Newtonian fluids, as will be discussed later on. First, however, we show that the viscosity is indeed changing significantly throughout the domain by looking at its probability distribution function in Figure~\ref{fig:visc}$b$, with Table~\ref{tab:cases} reporting the first three moments of the distributions. At a fixed Reynolds number, the viscosity spans a monotonically increasing range of values as $n$ is reduced, while the distribution shrinks when the Reynolds number is reduced at constant $n$. When $n \to 1$, the mean viscosity $\langle \mu \rangle$ tends to the classical Newtonian viscosity. Also, the distributions never reach the values of the infinite viscosity $\mu_\infty$, and very rarely approaches the zero one $\mu_0$; 
we can thus conclude that what observed hereafter is not influenced by the value of $\mu_0$ and $\mu_\infty$, which is true for large $\mu_0/\mu_\infty$.  It is thus representative of the broader family of power-law fluids. In the opposite limit when $\mu_0/\mu_\infty$ is tending to unity, the change of viscosity is rather limited, whatever is the value of $n$, and thus the Newtonian results are expected to hold again.

\begin{figure}
  \centering
  \includegraphics[width=0.49\textwidth]{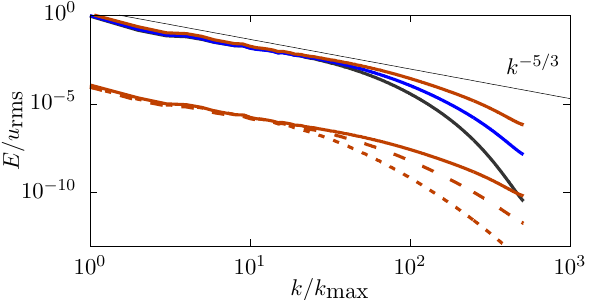}
  \includegraphics[width=0.49\textwidth]{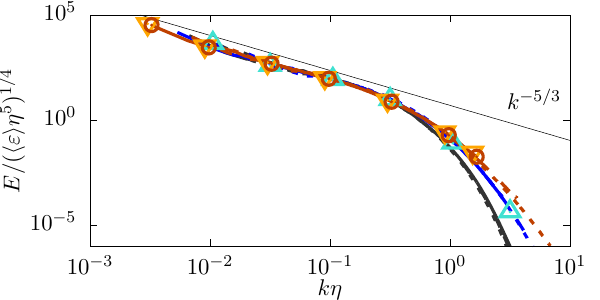}
  \caption{Energy spectra ($a$) for different power index $n$ at $Re_\lambda \approx 300$ (different colors), and for different Reynolds number $Re_\lambda$ with $n=0.4$ (different linestyles). The latter have been shifted downwards for clarity. ($b$) Energy spectra for different Reynolds number and power index, normalised with the Kolmogorov scale $\eta$ and turbulent dissipation rate $\varepsilon$.}
  \label{fig:spectra}
\end{figure}

The variable viscosity profoundly alters the flow, especially the velocity gradients, see e.g.~Figure~\ref{fig:visualization}. To understand how the variable viscosity is affecting the flow at all scales, we start by looking at the energy spectra $E(k)$, reported in Figure~\ref{fig:spectra}$a$ for different power index $n$ and different Reynolds numbers. For every power index $n$, the spectra exhibit an inertial range where the energy decays according to the classical Kolmogorov scaling $k^{-5/3}$, followed by a sharp decrease in energy within the dissipation range. As $n$ is reduced, we observe an extension of the inertial range of scales, and a shallower decay of energy at the smaller scales. Instead, as $Re_\lambda$ is reduced the typical reduction of the inertial range and expansion of the dissipation range are observed. Thus, the variable viscosity keeps unaltered the large scales of the flow, and brings higher level of energy at the small scales, with an extension of the inertial range and a weaker energy decay in the dissipation range.

While finding the Kolmogorov scaling in the inertial range of scales of a variable viscosity fluid may seem strange, it's appearance can be understood by rethinking the Kolmogorov hypotheses in our framework of variable viscosity. The hypothesis of local isotropy at the smallest scale remains unchanged, while the first and second similarity hypothesis need to be revisited. Indeed, both involve the definition of a viscosity and dissipation rate, which are then used to define the Kolmogorov scales, which are uniquely defined in classical turbulence while not so in our case of variable viscosity. As done before when computing the Taylor Reynolds number in Table~\ref{tab:cases}, here we take the mean viscosity $\langle \mu \rangle$ and the mean turbulent dissipation rate defined as $\langle \varepsilon \rangle = \langle 2 \nu \mathcal{S}_{ij} \mathcal{S}_{ij} \rangle$. With these two quantities, we can then extend Kolmogorov's theory in a straightforward manner, leading to the definition of the Kolmogorov scale $\eta=\left( \langle \mu \rangle^3/\langle \varepsilon \rangle \right)^{1/4}$, which allows to separate the small scales of motion $r \ll \eta$, where the statistics are supposed to be universal function of $\langle \mu \rangle$ and $\langle \varepsilon \rangle$ (first similarity hypothesis), from the intermediate ones $r \gg \eta$, where the statistics are independent of $\langle \mu \rangle$ and uniquely determined by $\langle \varepsilon \rangle$ (second similarity hypothesis). By simple dimensional argument, we can then rescale the energy spectra as done in Figure~\ref{fig:spectra}$b$, where we confirm that all the cases with different $n$ and $Re_\lambda$ indeed collapse into a master curve in the inertial range of scales. However, this is not the case in the dissipative region, where the spectra remain segregated by the different $n$, while collapsing for the different $Re_\lambda$. This is because in our case of variable viscosity the smallest scales should be expected to be universal function of $\langle \mu \rangle$, $\langle \varepsilon \rangle$, and $n$, being the latter an additional non-dimensional parameter; the same does not apply for the intermediate range of scales, where the effect of viscosity are not relevant. Thus, we can expect the Kolmogorov theory to hold true at the large and intermediate scales for $r\gg \eta$, while at the smallest scale a dependence on $n$ can still remain. In the following we will thus verify this, and study the dependence of the small scales from $n$.

\begin{figure}
  \centering
  \includegraphics[width=0.49\textwidth]{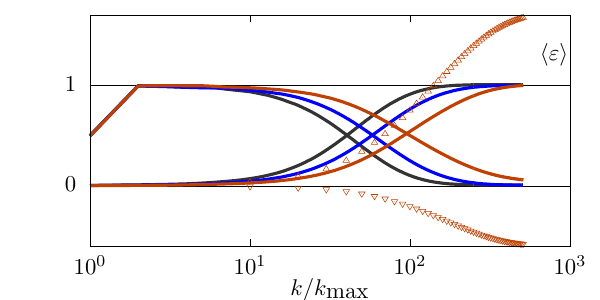}
  \includegraphics[width=0.49\textwidth]{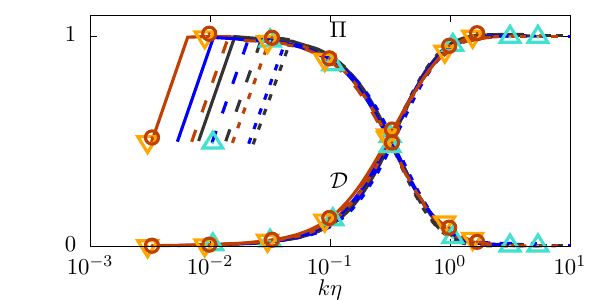}
  \caption{($a$) Energy balance for different power index $n$ at $Re_\lambda \approx 300$. For $n=0.4$, we also show the dissipation term decomposed into a mean ($\triangle$) and fluctuating ($\triangledown$) parts. ($b$) Energy balance for different Reynolds number and power index, normalised with the Kolmogorv scale $\eta$. The ordinates of both panels are divided by $\langle \varepsilon \rangle$.}
  \label{fig:balance}
\end{figure}

To gain a more detailed insight, we look at the scale-by-scale energy transfer balance
\begin{equation}
\mathcal{P} \left( k \right) + \Pi \left( k \right) + \mathcal{D} \left( k \right) = \langle \varepsilon \rangle,
\end{equation}
where $\mathcal{P} \left( k \right)$ is the production term associated with the external forcing, $\Pi \left( k \right)$ is the energy flux associated with the nonlinear convective term, and $\mathcal{D} \left( k \right)$ is the viscous dissipation, defined as $\mathcal{P} \left( \kappa \right) = \int_\kappa^\infty \left( \hat{f} \cdot \hat{u}^\star + \hat{f}^\star \cdot \hat{u} \right)/2 d\kappa$, $\Pi \left( \kappa \right) = \int_\kappa^\infty -\left( \hat{G} \cdot \hat{u}^\star + \hat{G}^\star \cdot \hat{u} \right)/2 d\kappa$, and $\mathcal{D} \left( \kappa \right) = \int_0^\kappa \left( \hat{D} \cdot \hat{u}^\star + \hat{D}^\star \cdot \hat{u} \right) d\kappa$. Here, $\hat{\cdot}$ denotes the Fourier transform operator, the superscript $\cdot^\star$ the complex conjugate, and $\hat{G}$ and $\hat{D}$ are the Fourier transform of the nonlinear and viscous terms \citep{pope_2001a}. Note that, the production term and the flux are obtained by integrating from $\kappa$ to $\infty$, while the dissipation term is integrated from $0$ to $\kappa$ to obtain a positive quantity that matches $\langle \varepsilon \rangle$; also, due to the variable viscosity, the usual relation $\hat{D} \left( \kappa \right) = 2 \nu k^2 E \left( \kappa \right)$ does not hold anymore.

In Figure~\ref{fig:balance}$a$, we show the energy fluxes and dissipation for different power index $n$ at large Reynolds number. The energy is injected into the system by the forcing trough $\mathcal{P} \left( k \right)$ which is not null only at the largest scale $k=k_\textrm{max}$ (not shown); energy is then carried from large to small scales by the advective term $\Pi \left(k \right)$, where it is ultimately dissipated by the viscous term $\mathcal{D} \left( k \right)$. The picture remains qualitatively the same when $n$ is reduced, except for a stronger advective term $\Pi \left(k \right)$ which dominates over a larger range of scales, confining the dissipation to even smaller scales. Here $\mathcal{D} \left(k \right)$ is defined with the local viscosity; we can decompose this term into a term coming from the mean viscosity $\langle \mu \rangle$ ($\triangle$), which is formally the same as in classical turbulence, and the rest which arises from its fluctuation $\mu - \langle \mu \rangle$ ($\triangledown$). The decomposition suggests that, the fluctuations of viscosity provide a negative dissipation, or in another words are a production mechanism of turbulent kinetic energy acting at the small scales, which is balanced on average by the excess of dissipation coming from the mean viscosity. 
Finally, Figure~\ref{fig:balance}$b$ shows again the energy balance for all $n$ and $Re_\lambda$, normalised with the Kolmogorov scaling. An excellent collapse of the curves is found. 

\begin{figure}
  \centering
  \includegraphics[width=0.49\textwidth]{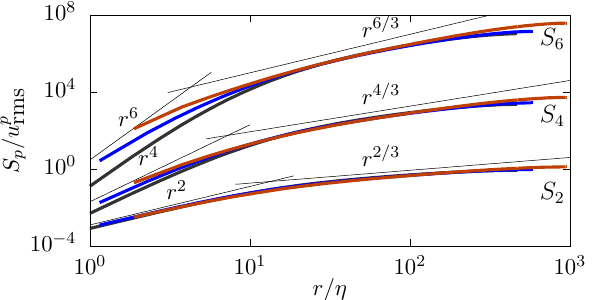}
  \includegraphics[width=0.49\textwidth]{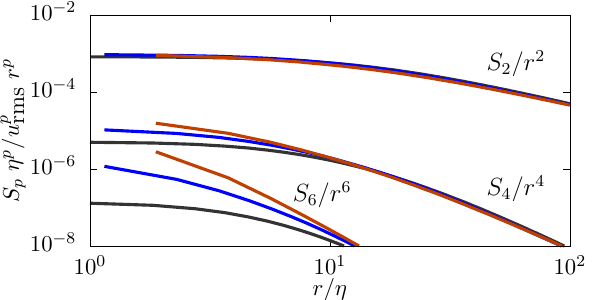}
  \includegraphics[width=0.49\textwidth]{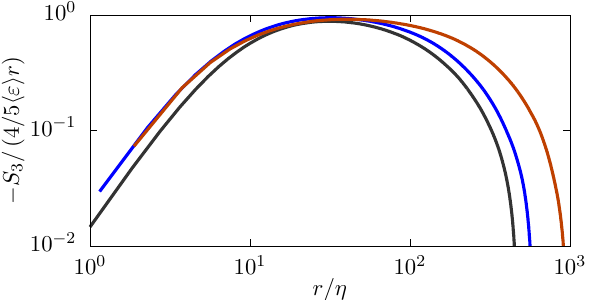}
  \includegraphics[width=0.49\textwidth]{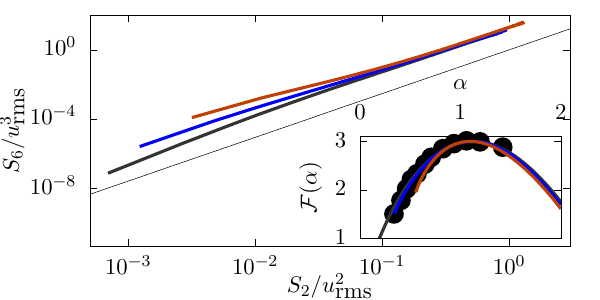}
  \caption{($a$) Structure function $S_p \left( r \right)$ for different power index $n$ at $Re_\lambda \approx 320$. The structure functions of orders $2$, $4$ and $6$ are shifted vertically for visual clarity. ($b$) Structure function $S_p \left( r \right)$ compensated with the expected scaling for $r \rightarrow 0$. ($c$) Compensated third order structure function $S_3 \left( r \right)$. ($d$) Extended self-similarity of structure functions, with the prediction based on the refined similarity hypothesis. The inset shows the multifractal spectra of the dissipation rate, with the symbols representing experimental data for a Newtonian fluid, taken from \citet{meneveau_sreenivasan_1991a}.}
  \label{fig:structure}
\end{figure}

The changes observed in the energy spectra can be appreciated also when looking at the structure functions in the real space; the longitudinal structure functions are defined as $S_p \left( r \right) = \langle \left[ \delta u \left( r \right) \right]^p \rangle $, where $p$ is the order of the moment and $\delta u \left( r \right) = u \left( x+r \right) - u \left( x \right)$ is the velocity increment across a length $r$. For the single-phase case, $S_p \sim r^p$ at small scales and $S_p \sim r^{p/3}$ in the inertial range, as predicted by the Kolmogorov theory \citep{kolmogorov_1941a}, with some deviations due to the flow intermittency \citep{ishihara_gotoh_kaneda_2009a}, caused by extreme events which are localised in space and time that break the Kolmogorov similarity hypothesis. These extreme events correspond to the large tails in the velocity increment distributions and make significant contribution to the high-order moments.

Figure~\ref{fig:structure}$a$ shows the structure functions of order $2$, $4$, and $6$ versus the separation $r$, for different value of the power index $n$ at large Reynolds number. We observe that the structure functions well collapse for different $n$ in the inertial range of scale, while they fan out at smaller scales in the dissipation range, consistently with the observation from the spectra. 
Assuming that the velocity is a differentiable function, we can expect $S_p \left( r \right) \sim r^p$ for $r\rightarrow 0$, and following \citet{schumacher_sreenivasan_yakhot_2007a}, we plot in Figure~\ref{fig:structure}$b$ the same structure functions, compensated with their expected scaling for the viscous range $r^p$, such that, the function must approach a constant for reducing $r$; this is evident for all the Newtonian structure functions, while for the shear-thinning fluid it is clear only for $S_2$ and less for the other orders, indicating that even smaller $r$ are needed to achieve the predicted scalings compared to a Newtonian fluid. Next, we assess the validity of the most important relationship valid in classical turbulence, the celebrated $4/5$-th law by Kolmogorov: $S_3 \left( r \right) = -4/5 \langle \varepsilon \rangle r$. Verifying its validity is the most direct proof of the possibility of extending Kolmogorov theory to shear-thinning fluids and on the correctness of using $\langle \varepsilon \rangle$. The results of our simulations are shown in Figure~\ref{fig:structure}$c$, where the third order structure function has been compensated with the expected value. The expected power law and the proper negative sign, which is connected to a direct energy cascade, are both visible, with an extension of the range where the scaling holds with $n$. Finally, we assess the flow intermittency using the extended self-similarity form \citep{benzi_ciliberto_tripiccione_baudet_massaioli_succi_1993a}, by plotting $S_6$ against $S_2$, as shown in Figure~\ref{fig:structure}$d$. In the limit case, where extreme events do not occur, the $S_6 \sim S_2^{6/2}$ power law holds, while the deviation from this behavior is a measure of the flow intermittency. It should be noted that there are already deviations from the theoretical prediction in the single-phase case, with the correction suggested by \citet{kolmogorov_1962a} offering a good prediction of the data. Moving to the effect of the variable viscosity, Figure~\ref{fig:structure}$d$ shows that a reduction of the power-law index generally leads to a larger deviation from the expected scaling at the small scales; the deviation from the single-phase is monotonic with $n$, and starts at larger $r$ as $n$ decreases, leading to a stronger flow intermittency. 


Examining the intermittency of the dissipation in space is an alternative method of examining the impact of extreme events. Using the procedure outlined by \cite{frisch_1995k}, we average the dissipation within a spherical region of radius $l$ to get $\varepsilon_l$. We pick a range of moments $-6 \le q \le 6$, $\varepsilon_l^p$, and average them over various locations and times. At large $l$, $\varepsilon_l$ is by definition equivalent to the bulk value $\langle \varepsilon \rangle$, while the two differs for small $l$, due to the localized regions of high dissipation in the fluid.
We assume that $\langle \varepsilon_l^q \rangle \sim l^{\tau_q}$ and then compute $\tau_q$ through fitting; finally, we obtain the multifractal spectra $\mathcal{F} \left( \alpha \right)$ using a Legendre transformation: $\alpha = d\tau / dq +1$ and $\mathcal{F} = q \left( \alpha -1 \right) - \tau_q +3$. The inset of Figure~\ref{fig:structure}$d$ shows the multifractal spectra for all cases at large Reynolds, and we observe that, all the multifractal spectra have peaks at $\alpha \approx 1$ and $\mathcal{F} \approx 3$, showing a background of space-filling dissipation; however, the presence of tails in the spectra indicates that the dissipation fields are not self-similar. No significant difference is found for the various $n$, indicating that the added intermittency observed in the velocity field is compensated by the variable viscosity in the dissipation rate.

\begin{figure}
  \centering
  \includegraphics[width=0.49\textwidth]{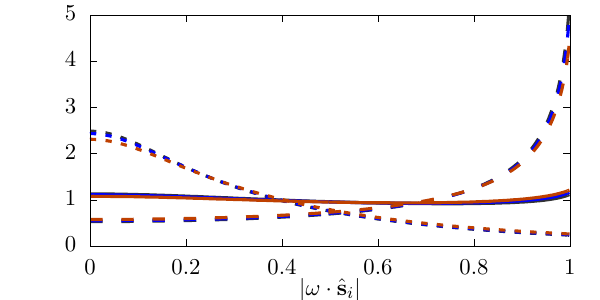}
  \includegraphics[width=0.49\textwidth]{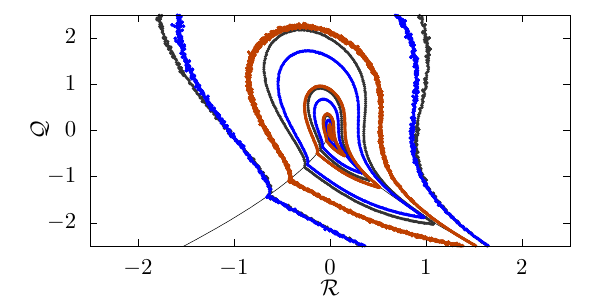}
  \caption{$(a)$ Histograms of the alignment of the vorticity unit vector $\hat{\boldsymbol{\omega}}$ with the eigenvectors (solid line) $\hat{\mathbf{s_1}}$, (dashed line) $\hat{\mathbf{s}_2}$, and (dotted line) $\hat{\mathbf{s}_3}$ of the strain-rate tensor. $(b)$ Joint histograms of $\mathcal{Q}$ and $\mathcal{R}$. The black curves show where the discriminant of the polynomial equation is zero, i.e., $27 \mathcal{R}^2/4 + \mathcal{Q}^3 = 0$. Data for $Re_\lambda \approx 300$.}
  \label{fig:local}
\end{figure}

To look more closely at the local flow altered by the variable viscosity, we compute the alignment of the unit-length eigenvectors $\hat{\mathbf{s}}_i$ of the strain-rate tensor $\mathcal{S}_{ij}$ with the vorticity $\boldsymbol{\omega} = \epsilon_{ijk} \partial u_k/\partial x_j \hat{\mathbf{e}}_i$ \citep{ashurst_kerstein_kerr_gibson_1987a}, where $\hat{\mathbf{e}}_i$ are the Cartesian unit vectors and $\epsilon_{ijk}$ is the Levi-Civita symbol. For an incompressible fluid, the three eigenvalues sum to zero, so the largest eigenvalue is never negative and its eigenvector $\hat{\mathbf{s}}_1$ corresponds to the direction of elongation in the flow. Similarly, the smallest eigenvalue is negative, and its eigenvector $\hat{\mathbf{s}}_3$ corresponds to the direction of compression in the flow. Finally, $\hat{\mathbf{s}}_2$ completes the tri-orthogonal system. Figure~\ref{fig:local}$a$ shows the probability-density functions of the the cosine of the angle between the vorticity $\hat{\boldsymbol{\omega}}$ and the eigenvectors $\hat{\mathbf{s}}_i$. There is a strong alignment of vorticity with the intermediate eigenvector $\hat{\mathbf{s}}_2$, which is frequently attributed to the axial stretching of vortices~\citep{ashurst_kerstein_kerr_gibson_1987a}, with the maximum probability at $|\hat{\boldsymbol{\omega}} \cdot \hat{\mathbf{s}}_2 | = 1$; the third eigenvector is largely perpendicular to the vorticity, producing a peak at $\hat{\boldsymbol{\omega}} \cdot \hat{\mathbf{s}}_3 = 0$, while the first eigenvector $\hat{\mathbf{s}}_1$ exhibits very little correlation with $\hat{\boldsymbol{\omega}}$. When the fluid is shear-thinning and $n$ is reduced, the same pictures remain, with an extremely weak tendency of vorticity aligning even more with the intermediate eigenvector $\hat{\mathbf{s}}_2$, and becoming more perpendicular to the last eigenvector $\hat{\mathbf{s}}_3$.

The three principle invariants of the velocity gradient tensor $\partial u_j/\partial x_i$ can be used to fully characterize the local topology of a flow~\citep{cheng_1996a}. The first invariant is zero due to incompressibility, the second invariant $\mathcal{Q} = \frac{1}{4}\omega_i\omega_i-\frac{1}{2}\mathcal{S}_{ij}\mathcal{S}_{ij}$ represents the balance between strain and vorticity, and the third invariant $\mathcal{R} = \frac{1}{4}\omega_i \mathcal{S}_{i j} \omega_j-\frac{1}{3} \mathcal{S}_{i j} \mathcal{S}_{j k} \mathcal{S}_{k i}$ represents the balance between strain and vorticity production. The roots of the polynomial equation $\Lambda^3 + \mathcal{Q} \Lambda + \mathcal{R} = 0$ are the eigenvalues of $\partial u_j/\partial x_i$. Figure~\ref{fig:local}$b$ shows the joint probability distributions for all of our flows at large Reynolds number. We also plot a line where the discriminant is zero: below this line, all three eigenvalues are real, and strain dominates the flow, whereas above this line, $\partial u_j/\partial x_i$ has one real and two complex eigenvalues, and vortices dominate the flow. The top-left and bottom-right quadrants feature tails in the distribution, which represent stretched vortices and areas where the flow compresses along a single axis~\citep{cheng_1996a}. The shape remains overall unchanged for the non-Newtonian fluids, but with the probability falling off much sharper when $n$ is reduced, thus indicating weaker velocity gradients, compensated by the local variation of the viscosity.

\section{Conclusions}\label{sec:conclusions}
In this work we have extended the Kolmogorov theory to shear-thinning fluids with variable viscosity, with the main quantities emerging being $\langle \mu \rangle$, $\langle \varepsilon \rangle$, and $n$, and with the Kolmogorov theory for a Newtonian fluid recovered when $n=1$. By analysing data from direct numerical simulations, we have shown that the variable viscosity brings no significant changes to the large and intermediate scales of the flow, while affecting only the smallest one. We have shown that these fluids exhibit a universal inertial range with a power-law spectrum scaling of $k^{-5/3}$, extending in range as the level of shear-thinning is increased, and caused by a strengthening of the nonlinear energy flux. As a consequence, the dissipation range shrinks. Remarkably, small scale universality is lost, with a remaining dependency on the power-law index $n$. The modified flows exhibit enhanced intermittency in the velocity field, with local flow structures having similar directionality to the Newtonian case, but with weaker intensity, the two effects being compensated by the variation of the viscosity. The results hold for different $\mu_0/\mu_\infty$, $\mathcal{K}$, and $n$, as long as the viscosity falls in the purely power-law region of the rheology. In particular, the effect of $\mathcal{K}$, while significant in changing the viscosity quantitatively, can be collapsed into a change in the actual Reynolds number experienced by the flow.

The results discussed in this work are relevant when studying polymeric flows, where often elasticity and shear-thinning are not decoupled. Since shear-thinning promotes the non-linear energy flux, while viscoelasticity weakens it, complex effects can arise when studying polymeric flows at both large and low Reynolds numbers: \textit{i)} shear-thinning acts against elasticity at large Reynolds numbers, potentially hiding the elasto-inertial range of scales \citep{rosti_perlekar_mitra_2023a}; \textit{ii)} shear-thinning can promote instabilities at low-Reynolds numbers \citep[see e.g. for a jet][]{soligo_rosti_2023a}, which are however not elastic, with the resulting turbulent flow which may be inertia-dominated, notwithstanding the low-Reynolds number \citep{amor_soligo_mazzino_rosti_2024a}. 

\FloatBarrier
\vspace{-0.5cm}
\small
\section*{Acknowledgments}
The author acknowledges the computer time provided by the Scientific Computing section of Research Support Division at OIST and by HPCI, under the Research Project grants \textit{hp210269}, \textit{hp220099}, \textit{hp230018}, \textit{hp250021} and \textit{hp250035}.

\section*{Funding} 
The research was supported by the Okinawa Institute of Science and Technology Graduate University (OIST) with subsidy funding to M.E.R. from the Cabinet Office, Government of Japan. M.E.R. acknowledges funding from the Japan Society for the Promotion of Science (JSPS), grant 24K17210 and 24K00810.

\section*{Declaration of Interests} 
The authors report no conflict of interest.

\normalsize
\bibliographystyle{jfm}
\bibliography{bibliography.bib}

\end{document}